\documentclass[]{aa}
\usepackage{natbib}
\usepackage{psfig}
\def\EBV{\mbox{E$_{\rm B-V}$}}

\def\Nmu{\mbox{N$_\mu$}}
\def\AV{\mbox{A$_{\rm V}$}}
\def\AB{\mbox{A$_{\rm B}$}}
\def\HH{\mbox{H$_2$}}
\def\XCH{\mbox{X$_{\rm CH}$}}         
\def\nH2{{\rm n}({\rm H}_2)}
\def\NH2{{\rm N}({\rm H}_2)}
\def\pccc{~{\rm cm}^{-3}} \def\pcc{~{\rm cm}^{-2}}

\def\Tstar#1 {\mbox{${\rm T}_{\rm #1}^*$}}
\def\Tsub#1 {\mbox{$T_{\rm #1}$}}
\def\TK  {\Tsub K }

\def\Tbg  {\Tsub bg }

\def\TA  {\Tsub A }
\def\Texc {\Tsub exc }

\def\Tcmb{\Tsub cmb }

 \def\arcmin{\mbox{$^{\prime}$}}
\def\omet{\mbox{$(1-{\rm e}^{-\tau})$}}

\def\degr{$^{\rm o}$}
\def\p{$^+$}

\def\hcop{\mbox{{HCO\p}}}

\def\cch{\mbox{C$_2$H}}
  
\def\hhco{\mbox{H$_2$CO}}
\def\h13cop{\mbox{{H$^{13}$CO\p}}}

\def\c3h2{\mbox{C$_3$H$_2$}}
 
 \def\R0{R$_0$}

\def\ddeg{{}^\circ\kern-.1em}

\def\kms{\mbox{km\,s$^{-1}$}}
\def\ps{\mbox{s$^{-1}$}}
\def\bll{BL Lac}

\def\E#1 {$10^{#1}$}
\def\E#1 {E{#1}}
\def\P#1,{$\nH2\TK~=~#1\times~10^4\pccc$~K}
\def\ec#1,#2,#3,{#1\,(#2)\E{#3}}

\def\zoph{$\zeta$ Oph}

\def\H3{\mbox{H$_3$}}

\def\ammon{\mbox{N\H3} }

\title{Comparative Chemistry of Diffuse Clouds IV: CH}

\author{H. Liszt\inst{1}\ and R. Lucas\inst{2}}
\institute{National Radio Astronomy Observatory,
           520 Edgemont Road,
           Charlottesville, VA,
           USA 22903-2475
\and       Institut de Radioastronomie Millim\'etrique,
           300 Rue de la Piscine,
           F-38406 Saint Martin d'H\`eres,
           France}

\begin{document}
\date{received \today}
\offprints{H. S. Liszt}
\mail{hliszt@nrao.edu}

\abstract{
We observed the 3335 MHz ($\lambda$ 9cm) F=1-1 line of CH toward a 
sample of diffuse clouds occulting compact extragalactic mm-wave 
continuum sources, using the old NRAO 43m telescope.  Because 
radiofrequency observations of CH really must be calibrated with 
reference to a known CH abundance, we begin by deriving 
the relationships between CH, \EBV, \HH\ and other hydrides found by
optical spectroscopy. No simple relationship exists between 
N(CH) and \EBV, since N(CH) is strongly bimodal with respect to
reddening for \EBV $<$ 0.3 mag and the typical range in the N(CH)/\EBV\ ratio 
is an order of magnitude or more at any given \EBV $> 0.3$ mag.  However, 
N(CH)/N(\HH)  $=   4.3 \pm 1.9 ~\times~10^{-8}$ in the mean and
N(CH) $\propto$ N(\HH)$^ 1.00\pm0.06$ for 
$10^{19} < $ N(\HH) $ < 10^{21}\pcc $.  If CH is a good predictor of \HH, 
40\%-45\% of the hydrogen in the local diffuse/translucent ISM is in the 
molecular form at the accepted mean density, higher than previous estimates 
found in samples of lower-than-average mean density.  
Optical observations of the population ratios in
the upper and lower halves of the CH lambda-doublet suggest 
that the brightness of the 3335 MHz CH line should be double-valued at a 
given CH column density in diffuse gas:  double-valuedness
is noticeable in our data when comparing CH with CO or \hcop. The CH 
brightness at 3335 MHz is mildly bimodal with respect to CO emission 
in our diffuse cloud data but much more strongly bimodal when comparing 
diffuse or translucent gas and dark gas.  The CH $\Lambda$-doublet is 
generally inverted in diffuse gas but we did not succeed in measuring the 
excitation temperature except toward 3C123 where we confirm one older 
value $\Texc \approx -10$ K.
}
\maketitle

\section {Introduction.}

\begin{figure*}
\psfig{figure=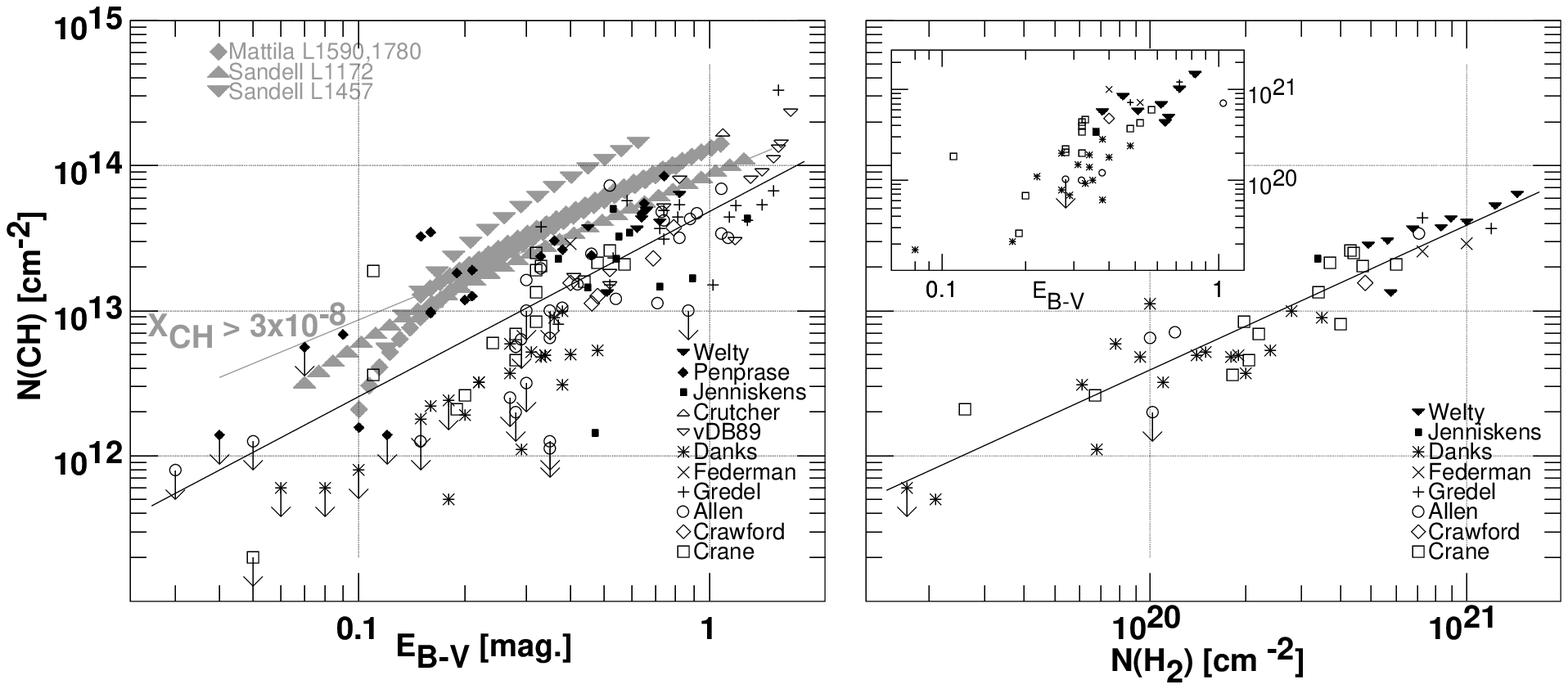,height=7cm}
\caption[]{The dependence of optically-measured CH column densities 
on reddening (left panel) and N(\HH).  CH Data are from  \cite{CraLam+95},
\cite{Cra95},  \cite{All94}, \cite{GreVan+93}, \cite{FedStr+94}, 
\cite{DanFed+84}, \cite{vDB89}, \cite{Cru85}, \cite{JenEhr+92},
\cite{Pen93}  and Welty et al. as tabulated in \cite{RacSno+02} .  
The \HH\ column densities are from \cite{SavDra+77} and \cite{RacSno+02} 
except for one line of sight each from \cite{JosSno+86}, 
\cite{SnoRac+00} and \cite{RacSno+01}.
Superposed in the left panel are the N(CH)-\EBV\ loci derived from the 
dark cloud studies of \cite{Mat86} for L1590 and L1780, and \cite{San82} for 
L1172 and L1457;  also shown there is the line corresponding to 
\XCH\ $\approx$ 2N(CH)/N(H) $ = 3\times10^{-8}$ as discussed in the text.  
Superposed in the right-hand panel are a plot of N(\HH) {\it vs.} \EBV\ for
sightlines with measured N(CH) and the best-fit power-law regression line
(see Sect 2.2).}
\end{figure*}

Some chemical species present in diffuse interstellar gas can only be 
studied in  optical/uv absorption:  these include
C$_2$ \citep{ChaLut78}, NH \citep{MeyRot91,CraWil97}, HCl \citep{FedCar+95},
and most importantly \HH\ \citep{SavDra+77}.
Many more, chiefly polyatomics and heavy diatomics, are seen only in the 
radio regime.  These include
\hhco\ \citep{FedWil82,LisLuc95a}, 
\c3h2\ \citep{CoxGue+88,LucLis00a},
\ammon\ \citep{Nas90}, 
\hcop\ \citep{LisLuc96,LisLuc00},
\cch\ \citep{LucLis00a},
HCN and HNC \citep{LisLuc01},
SiO \citep{LucLis00}, and sulfur-bearing species such as  
CS, SO, H$_2$S, HCS\p\ $etc.$ \citep{LucLis02}.

The few species which overlap both domains (CH, OH, CN and CO) provide an 
important  bridge between two rather different ways of studying 
interstellar chemistry in diffuse clouds.  CN, for instance, whose 
relationships with CH, C$_2$, CO and \HH\ are well-studied optically  
\citep{FedStr+94,vDB89,FedLam88}, is closely tied to HCN and HNC and 
more loosely related to perhaps a dozen other polyatomic species seen 
at mm-wavelengths \citep{LisLuc01}.  
Similarly OH, whose abundance relative to \HH\ was 
found by \cite{Cru79a} to be constant over a wide range of extinction, 
is closely tied to \hcop\ \citep{LisLuc96,LucLis96,LisLuc00}, leading us 
to believe that the relative abundance of \hcop\ also varies little.  
\hcop\ is the immediate progenitor of CO in diffuse gas 
\citep{GlaLan75,GlaLan76,BlaVan86}.  Its unexpectedly large abundance 
relative to  \HH\ is sufficient to explain the run of N(CO) with N(\HH) for 
$2\times 10^{19} < $ N(\HH) $< 2 \times 10^{21} \pcc$.
$4\times 10^{12} < $ N(CO) $< 2 \times 10^{16} \pcc $ \citep{LisLuc00}.

CH, the subject of this study, represents another important bridge
between the optical and radio regimes, all the more so because its
abundance relative to molecular hydrogen is very well-determined and
nearly fixed for lines of sight having
\AV $\leq 3$ \citep{FedLam88,DanFed+84,Fed82}.  A determination of its
column density in the gas toward our compact extragalactic mm-wave 
continuum background sources would provide an important check on
the abundances of all the other species we detect.  Unfortunately,
observation of CH at 3335 MHz in the radio regime is presently impossible
and opportunities will be at best severely restricted in the future.
Of the large single dishes, only Arecibo and Nancay apparently have 
current plans to provide a receiver in the 9cm waveband.   Eventually, the 
complete frequency coverage of the EVLA will make possible widespread 
study of CH in stimulated emission against continuum sources, although 
the usefulness of such work, absent complementary emission 
studies, may be somewhat limited. 

In this work we present 3335 MHz F=1-1 CH observations in the direction
of  compact extragalactic continuum sources which we have previously
observed in an extensive, ongoing study of mm-wave absorption line 
chemistry. We took data toward and around these and two other sources 
which have stronger cm-wave continuum and are occulted by denser (but
not necessarily darker) gas  ($i.e.$ 3C123 and 3C133),
in much the same way as we did earlier for OH  \citep{LisLuc96,LucLis96} 
and \hhco\  \citep{LisLuc95a}.  We did so at least partly for the purpose 
of deriving the excitation temperature in diffuse gas and did not see 
any true CH absorption:  the optical depth was either negative, owing 
to the well-known collisionally-induced  inversion of the ground-state 
lambda-doublet \citep{BerChe+76,BujSal+84,BouRie+84}, or immeasurably small.  
The CH transition is almost certainly inverted in the diffuse gas 
we observed but excitation studies will require a telescope 
gain much larger and a telescope beam much smaller than that afforded 
by the old NRAO 43m antenna.  In the absence of much direct evidence
allowing a calibration of the relationship between N(CH) and the
microwave brightness in diffuse gas, we resort to a discussion of 
the rich body of optical CH measurements in the diffuse ISM.
 
The plan of the present paper is as follows.  In Sect. 2 we derive 
the abundance of CH with respect to \EBV, \HH\ and other hydrides, 
using optical  absorption line observations, and we note the implications 
of optical studies of the excitation of the ground-state CH 
$\lambda$-doublet for microwave observations.  In Sect. 3 we describe our 
new 9cm observations of the 3335 MHz CH F=1-1 (main) line, and compare 
them to our prior results for OH, \hcop, \cch\ and CO in the same 
directions. Sect. 4 is a brief summary.

\section{Optical absorption line studies of CH}

\subsection{The relationship between CH and \EBV}

Much of the original rationale for using CH as a tracer of \HH\ was the 
relationship N(CH)$ = 6.3\times10^{13}\pcc~\EBV$ derived by \cite{LanWil78} 
using a combination of optical and microwave determinations of N(CH) along 
some twenty lines of sight; for instance, see \cite{Mat86} which forms the 
basis of the discussion by \cite{MagOne95}.  The microwave 
determinations are subject to considerable uncertainty (at least 
comparatively) so that an optical determination of N(CH) is preferable, 
at least initially. \cite{Lie84} shows how to derive 
N(CH) properly from the available CH optical absorption lines.

Figure 1 at left shows the run of N(CH) with \EBV\ found in a much larger 
sample of more recent optical measurements
\citep{All94,CraLam+95,Cra95,FedStr+94,GreVan+93,DanFed+84,vDB89,Cru85,
JenEhr+92,Pen93} including the data of Welty et al. tabulated in
\cite{RacSno+02}.  Represented are 140 lines of sight harboring
 120 CH detections: many uninformative upper limits contained in the original 
references were not transcribed.  The most recent measurement was used in those 
cases where lines of sight had been observed more than once.  It is not 
possible to infer a single (or single-valued) or linear relationship between 
N(CH) and \EBV.

For $ \EBV < 0.3$, N(CH) is largely bimodal, i.e  either 
N(CH)$ \la 3\times10^{12}\pcc$, in, presumably, typical diffuse gas, 
or N(CH) $ > 10^{13}\pcc$ in the CO-emission-selected, high-latitude
objects studied by \cite{Pen93} and along one of the sightlines
studied by \cite{CraLam+95}.  In these directions, the gas is 
apparently dense enough or sufficiently poorly-illuminated for the 
great majority of hydrogen to have been converted into $\HH$ even 
for $\AV < 1$.  The transition to consistently high CH abundances occurs 
somewhere in the range $0.1 \le \EBV \le 0.4$; many lines of sight with 
\EBV\ = 0.3 mag do not show CH and one at \EBV\ = 0.87 from  \cite{All94} 
is underabundant compared to the mean by about a factor of 3.  

\begin{figure}
\psfig{figure=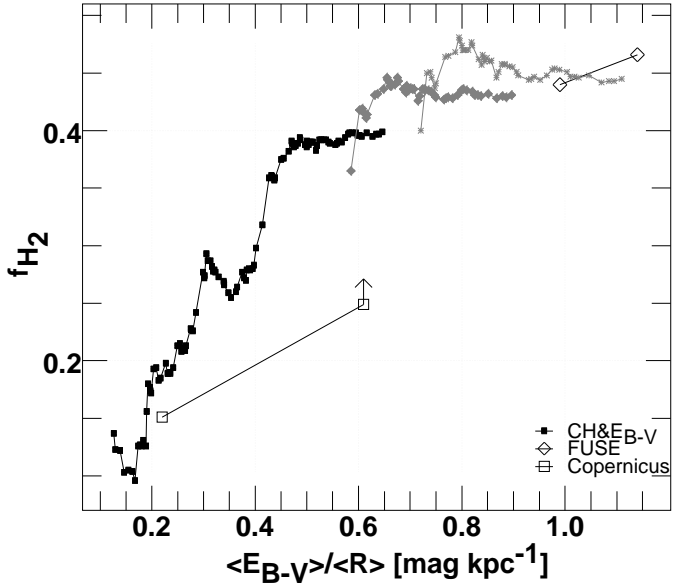,height=7.7cm}
\caption[]{The fraction of H-nuclei in \HH\ derived by various means.
The connected symbols labeled ``Copernicus'' represent the results of
\cite{BSD78}, measured (at left) and corrected for bias to lower
than average mean density.  \HH+CH' represents the mean over sightlines
with measured N(CH) and N(\HH) as shown in the inset to Fig. 1 at right,
assuming N(H) = $5.8\times10^{21}~\pcc$ \EBV.  The chained curves
represent samples of the sightlines with measured N(CH), assuming
the same N(H)-\EBV\ conversion and \XCH\ $=  4.3\times10^{-8}$ as described
in Sect. 2.2.1}
\end{figure}

Many of the points seem to lie
along a band increasing approximately as N(CH) $ \propto \EBV^{1.8}$ 
for $0.1 \leq \EBV \leq 1$; this arises because N(\HH) increases with 
\EBV\ and \XCH\ $\equiv$ N(CH)/N(\HH) is approximately constant 
(see Sect. 2.2). The enormous variance of the observed N(CH) values near 
\EBV\ = 0.3 mag seems quite extraordinary even if order-of-magnitude 
variations in N(CH) are common at most other \EBV\  as well.  Perhaps a 
second transition to yet more fully molecular gas will become evident 
in the N(\HH)-\EBV\ relationship as more data become available.

Shown in the left-hand panel of Fig. 1 is the line 
N(CH) $= 3\times10^{-8} \times 5.8\times10^{21}$ \EBV$ \pcc/2$,
the maximum CH column density which could arise if 
\XCH\ $= 3\times10^{-8}$ and all gas along the
line of sight were in the form of \HH\ ({\it i.e.} \EBV\ is converted
to a column density of H-nuclei using the standard value of \cite{BSD78}).
 Apparently, a few sightlines with reddening 
as small as 0.1-0.2 mag have \XCH\ approaching $10^{-7}$.

\begin{table*}
\caption[]{Column densities of hydrides in diffuse and dark gas}
{
\begin{tabular}{lcccccc}
\hline
         &  \zoph &    $\zeta$ Per & $o$ Per & HD27778 & TMC-1$^5$ & L134N$^5$ \\
\hline 
 & N($\pcc$) & N($\pcc$) & N($\pcc$) & N($\pcc$) & \Nmu($\pcc$) & \Nmu($\pcc$) \\
\hline
CH$^1$ & 2.50E13  & 2.03E13  & 1.90E13  & 2.90E13 & 2E14 & 1E14 \\
NH$^2$ & 8.8(1.2)E11  & 9.0(0.2)E11  &  & 2.7(0.6)E12 & & \\
OH$^3$ & 4.7(0.7)E13  & 4.1(0.4)E13  & 7.8 (2.6)E13 & 10.2(0.4)E13 & 3E15 & 7.5E14 \\
\HH$^4$ & 4.4E20  & 4.8E20  & 4.0E20  & 1.0(0.3)E21 & (1E22) & (1E22) \\

\hline
\end{tabular}}
\\
$^1$ N(CH) from \cite{CraLam+95} except HD27778 from \cite{FedStr+94} \\
$^2$ N(NH) from \cite{MeyRot91} except \zoph\ from \cite{CraWil97} \\
$^3$ N(OH) from \cite{Rou96} and \cite{FelRou96} \\
$^4$ N(\HH) from \cite{SavDra+77} except HD27778 from \cite{JosSno+86} \\
$^5$ Dark-cloud data from \cite{OhiIrv+92} \\
\end{table*}

Superposed in the left panel of Fig. 1 we have included the mean CH-\AB\
loci derived for several dark clouds from microwave measurements of the
3335 MHz line, using \AB\ = 4\EBV\ as in the
original references.  In what follows we will denote by \Nmu(CH) a 
CH column density derived by converting a microwave CH intensity
using standard formulae.  For comparison, then, we note that \cite{Mat86} 
used the 3335 MHz line of CH to derive 
\Nmu(CH) $= (13.9\pm1.2) \times10^{13}\pcc (\EBV-0.085)$
and \Nmu(CH)$ = (13.9\pm1.2) \times10^{13}\pcc (\EBV-0.06)$ in the two
dark clouds L1590 and L1780.  \cite{San82} found relationships  
\Nmu(CH) $= 27.2 \times10^{13}\pcc (\EBV-0.1)$ for L1457 and
\Nmu(CH) $= 9.2 \times10^{13}\pcc (\EBV-0.03)$ for L1172 while the much 
more limited data for L1642 in Table 4 of  \cite{SanJoh+81} have a 
limiting slope of \Nmu(CH)/\EBV $= 11.2 \times10^{13}\pcc$.  

There is a very substantial range in the limiting behavour of the 
derived \Nmu(CH)/\EBV\ ratios in these objects, and all are much higher 
than orignally derived in the work of \cite{LanWil78}.  Nonetheless,
they do in general correspond fairly well to the values 
of \XCH\ seen in the upper reaches of the CH column densities derived
optically.  The use of high N(CH)/\EBV\ ratios is probably acceptable,
even at low \EBV, for CO-emitting gas.

\subsection{The relationship between CH and \HH}

Much of the behaviour seen in the left-hand panel of Fig. 1 can be understood
simply on the basis of the observed variation of N(\HH) with reddening,
and a  fixed or nearly-fixed abundance of CH relative to 
\HH.  Such a nearly linear relationship between 
N(CH) and N(\HH) was demonstrated by \cite{Fed82} and by \cite{DanFed+84}
and relatively little had changed in the intervening time {\it vis-a-vis}
\HH\ until the recent FUSE survey work of \cite{RacSno+02}.  
The CH profiles have improved somewhat in quality and number and the important 
reanalysis by \cite{Lie84} clarified the interpretation greatly.  Fig. 1 at 
right shows the current situation, from which, for the lines of sight with
detections of both CH and \HH, we derive 
$\log{{\rm N(CH)}} = (-7.35\pm1.31) + (1.00\pm0.06)\log{N(\HH)}$. 
A less comprehensive version of this diagram which provided very nearly
the same regression line was shown by \cite{MagOne+98}.

In the mean,  \XCH\ $= 4.3 \pm 1.9\times10^{-8}$.
The $\pm45$\% variance of N(CH) about N(\HH) is appreciable; 
it sets the ultimate limit on expectations of the reliability of CH as 
a predictor of\HH. These data represent sightline averages over multiple 
components and do not prove that the CH abundance is constant in individual 
clouds of higher density and extinction. Indeed, the dark cloud data 
discussed in Sect. 2.4 show that \XCH\ declines markedly in darker gas 
and a steady decline from diffuse to dark conditions has been noted since 
the original microwave surveys of \cite{RydKol+76}, \cite{HjaSum+77}
and \cite{Mat86}.  The diminished scatter in Fig. 2 at right for the lines
of sight having the highest N($\HH$) is strongly suggestive of blending.

As shown in the inset in the right-hand panel of Fig. 1, much of the scatter 
in the plot 
of N(CH) {\it vs.} \EBV\ can be understood simply in terms of the
variation of N(\HH) with \EBV, given that \XCH\ is approximately constant.
Along lines of sight where \HH\ is known directly,  N(\HH) increases 
approximately as \EBV$^{1.8}$ for \EBV $> 0.2$.  The \cite{Pen93} sample,
selected on the basis of relatively strong CO J=1-0 emission, probably 
represents a special circumstance whereby the molecular fraction is high 
even below \EBV = 0.3.

\subsubsection{The fraction of gas in molecular form}

The fraction of gas in molecular form in the local ISM is an important 
quantity which provides interesting constraints on interpretations of 
atomic and molecular gas constituents independently.  Here we take a 
new approach to determining a local mean for 
f$_{\HH} ~\equiv~ 2$ N(\HH)/(N(H I) + 2N(\HH)), 
based on the near-constancy of \XCH\ and the existence of a substantial 
set of  CH measurements which sample a wider range of mean line-of-sight 
densities than has been directly observable in H I and \HH : a sample whose 
mean density is more nearly equal to that of the true mean in the local ISM 
offers the possibility that the mean molecular gas fraction might be more 
representative of the overall molecular fraction in the nearby ISM as well.  

Shown in Fig. 2 are the results of estimating $<{\rm f}_{\HH}>$ in three 
ways. The symbols labelled `Copernicus' are taken from \cite{BSD78} and 
represent their results along nearly 100 lines of sight of much lower than 
average mean density $<$\EBV$>$/$<$R$>$ = 0.22 mag kpc$^{-1}$. The accepted 
mean is 0.61 mag kpc$^{-1}$, from \cite{Spi78} based on the earlier discussion
of \cite{Mun52}
\footnote{Reddening per unit distance has units of density if
\EBV $\propto$ N(H); R is the distance to a background star. See
\cite{Spi78} }.
The symbol at lower mean density is the actual measurement; at higher 
mean density it is their estimate of the true value, corrected for bias.
The symbols labelled `FUSE' represent the new data of
\cite{RacSno+02}; the symbol at lower mean density
represents only those lines of sight for which N(H I) was not
estimated by assuming a proportionality to \EBV. 
The chained lines labelled 'CH\&\EBV ' represent samples of the lines 
of sight with CH measurements, assuming 
N(\HH) = N(CH)$/4.3\times10^{-8}$ and 
N(H) = N(H I)+ 2N(\HH) $= 5.8\times10^{21}~\pcc$ \EBV, so that
$<$N(\HH)$>/<$N(H)$> = 4.01\times^{-15}$ $<$N(CH)$>/<$\EBV$>$, as we 
now discuss.

To construct the CH-based samples, we sorted the lines of sight in order
of increasing individual \EBV/R  and  derived $<$N(\HH)$>/<$N(H)$>$
for contiguous, progressively larger samples of four or more sightlines 
beginning at different lower cutoffs. Where only upper limits on N(CH) 
were available, N(\HH) was taken to be zero.  So each chained line segment
in Fig. 2 shows 
the mean molecular fraction, derived under the idealized assumption of a 
constant CH abundance and gas-dust ratio, as the sample mean extinction per 
unit distance varies.  The dark, left-most line labelled ``CH\&\EBV'' begins
at the lowest observed \EBV/R and contains all lines of sight at its 
right-most extent; it just barely extends beyond
$<$\EBV$>/<{\rm R}>  = 0.61$ mag kpc$^{-1}$ when all the lines of sight are 
included.

The Copernicus and FUSE measurements (not the `corrected' Copernicus  
estimate) agree entirely with the CH-based samples; the high mean molecular 
fraction seen by FUSE should not be dismissed as biased because of the 
high sample mean density.  From the CH data we see that the mean 
molecular fraction appears to increase fairly rapidly at low sample mean 
densities and is of order 0.4 - 0.45 at the accepted local mean density.

\subsection{Population ratios}

\begin{figure}
\psfig{figure=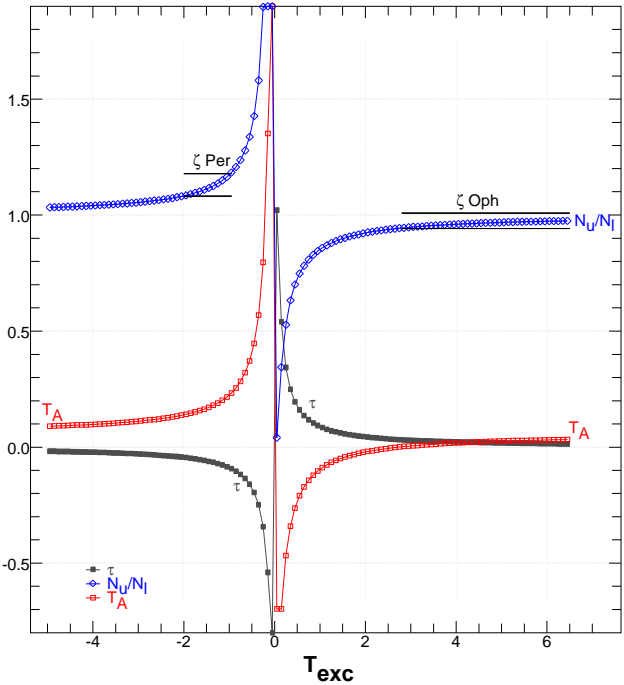,height=9cm}
\caption[]{Population of the CH $\Lambda$-doublet. Plotted as a 
function of the excitation temperature of the CH ground-state 
$\Lambda$-doublet are the population ratio of the upper and lower 
states N$_u$/N$_l$ (upper curve; chained diamonds) and the optical depth 
(shaded)
and antenna temperature of the 3335 MHz F=1-1 transition for 
N(CH)/$\Delta V$ = (N$_u$+N$_l)/ \Delta V = 2.5\times10^{13}\pcc(\kms)^{-1}$, 
$\eta_B = 0.7$.  The horizontal bars labelled ``$\zeta$ Per'' and 
``\zoph'' represent the $\pm1\sigma$ range of population ratios 
observed by \cite{JurMey85}.  See Sect. 2.3 of the text.}
\end{figure}

As noted in the Introduction, the excitation temperatures of the microwave 
CH transitions, needed to convert observed brightnesses to column density 
in diffuse gas, have not been (and probably cannot be) determined from 
microwave observations.  However,
 \cite{Lie84} pointed out that the populations of the upper and lower
levels of the ground-state CH $\Lambda$-doublet could be derived 
separately by observing different optical absorption lines (see his 
Fig. 2).  He derived the excitation temperature $\Texc$ from extant 
observations toward several well-studied stars and found that it 
could be either slightly negative or positive, but with rather 
large uncertainty.  Extant models of CH excitation by 
 \cite{BerChe+76} predicted a transition from normal excitation 
(\Texc\ $>$ 0 K) to substantial inversion (\Texc\ = -0.6K) at a density
somwehere between $10^2$ and $10^3 ~\HH\pccc$, so it was (and is)
reasonable to believe that both branches of the excitation conditions 
would manifest themselves in diffuse gas.
 
 \cite{JurMey85} rose to the challenge of taking data sufficiently
accurate for this purpose and their results are summarized in
Fig. 3 here.  The horizontal axis is the excitation temperature
of the $\Lambda$-doublet (the optical structure does not resolve
the hyperfine splitting). We define \Texc\ in the usual way
by asserting that the upper and lower halves of the doublet 
(which have equal statistical weights) are in the ratio 
$N_u/N_l = \exp(-h\nu/k\Texc)$ with $\nu$ = 3335 MHz, $h\nu/k = 0.160$K
corresponding to the so-called main (F=1-1) line.  We further assume that
the excitation temperature of the doublet as a whole applies to the 
microwave lines individually, and plot the antenna temperature and optical depth 
of the F=1-1 line assuming a beam efficiency of $\eta_B = 0.7$
and  N(CH)/$\Delta V$ = (N$_u$+N$_l$)/$\Delta 
V = 2.5\times10^{13}\pcc(\kms)^{-1}$, using the column density
observed toward \zoph\ (see Table 1; formulae sufficient to
reproduce this plot are given in Sect. 3.1). The plotted curves then represent 
the hypothetical antenna temperature and optical depth of a 1 
\kms-wide line, or the profile integral of either quantity if $|\tau|$
is small.

Around the curves of $N_u/N_l$ we have placed horizontal error bars 
displaced by $\pm1\sigma$ about the mean for the lines of sight toward 
$\zeta$ Per and $\zeta$ Oph.  \cite{JurMey85} found two components with 
similar conditions toward $\zeta$ Per; their results for $o$ Per 
(not represented here) are similar to those for $\zeta$ Oph.   For 
$\zeta$ Per, the error bars extend to meet the curve of population 
ratio and are meant to show the $\pm1\sigma$ error in \Texc. For 
\zoph\ the case cannot be summarized quite so concisely. The $-1\sigma$ 
error bar actually extends indefinitely to the right since the population 
ratio is bounded above at unity for $\Texc > 0$.  The $+1\sigma$ error bar 
for \zoph\ also has a branch to the far left (since it lies above unity 
in the population ratio) and population ratios above unity are consistent 
with the \zoph\ measurements at about the $1\sigma$ level.

\begin{figure*}
\psfig{figure=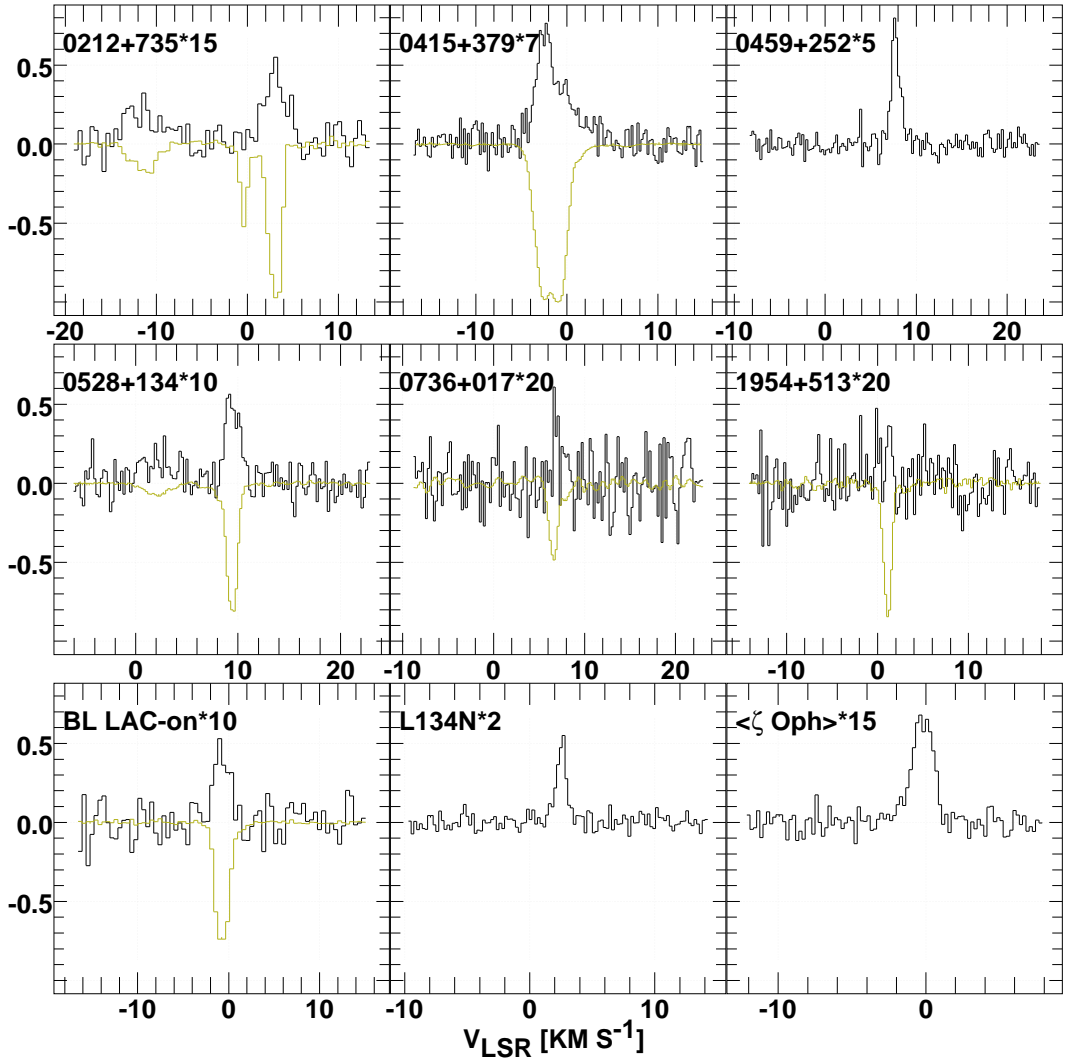,height=14cm}
\caption[]{CH emission profiles (units of Kelvins).  For the 
lines of sight around continuum sources, the profiles are averages 
of off-source spectra taken 11\arcmin\ North, East, South and West 
(except for \bll, which is on-source).  Shown shaded, where possible, 
is the corresponding 89 GHz \hcop\ J=1-0 absorption profile from 
\cite{LucLis96} in the form line/continuum-1.  At lower right is a 
spectrum averaged over 80\arcmin\ in a North-South direction around 
\zoph\ from \cite{Lis97}, and a comparison spectrum toward L134N. The 
emission spectra have been scaled in various ways, as indicated.}
\end{figure*}

\begin{figure*}
\psfig{figure=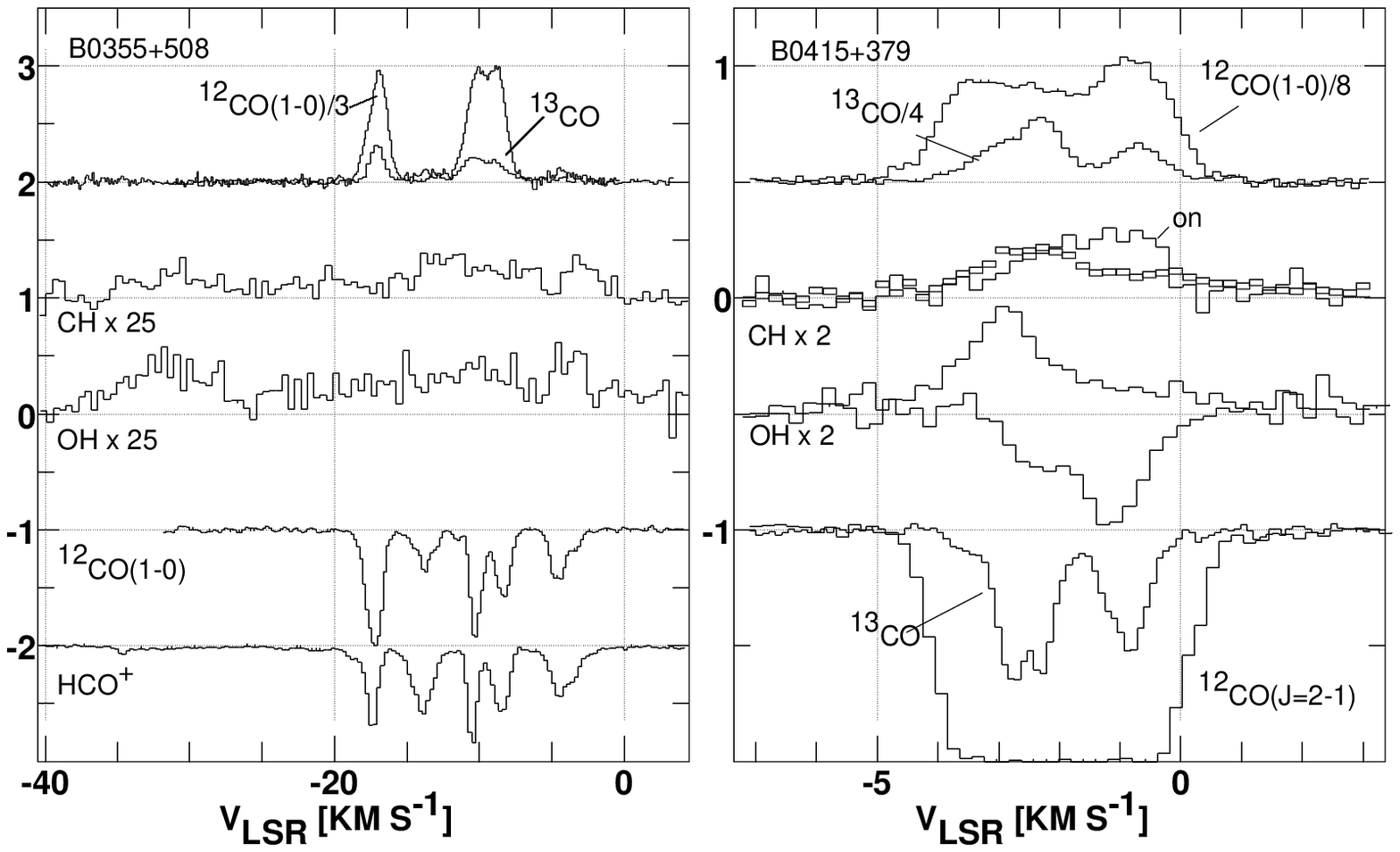,height=9cm}
\caption[]{Spectra of CH and other species toward B0355+508 (left)
and B0415+379.  For B0415+379 (3C111) both on-source and off-source
spectra are shown for CH and OH.  The on-source OH spectrum and all
the other obvious absorption spectrum are presented in the form
(line/continuum-1) while the on-source CH spectra and all the
obvious emission spectra are in K. At right, the CO emission 
spectra at top are for J=1-0 while the absorption spectra at bottom
are J=2-1.}
\end{figure*}

The optically-derived population ratios nominally predict that the 
microwave brightness should be much larger for $\zeta$ Per; a 
factor three is cited by \cite{JurMey85}, even given that N(CH) is 
25\% higher toward \zoph.  However, this is not observed to be
the case.  Both directions have main-line profile 
integrals (integrated antenna temperature)
of 0.08-0.09 K \kms\ if our results for 
$\zeta$ Oph (shown below and in \cite{Lis97}) can be compared to those 
of  \cite{HjaSum+77} or  \cite{Wil81} for $\zeta$ Per.  Unfortunately, this 
similarity was obscured for \cite{JurMey85} because \cite{Lie84} quoted a 
profile integral which was twice as large for $\zeta$ Per as for \zoph.

It appears that the 3335 MHz CH line toward both stars is
inverted, with (nominally) \Texc\ $\approx -5$ K for \zoph\ and 
\Texc\ $\approx - 3$ K for $\zeta$ Per, if the broad  microwave and narrow 
optical lines of sight sample approximately equal values of N(CH)
\footnote{that is, we may infer comparable excitation
from the radio data only if N(CH) is similar in the two directions, as
is the case along the narrow optical lines of sight; see also \cite{Wil81}}.  
These $|\Texc |$ are small enough that a noticeable  error occurs 
if \Nmu(CH) is derived in the limit $\Texc \rightarrow -\infty$;
N(CH) would be overestimated by 60-100\% .

In summary, even though the radio data do not show the most extreme 
differences in excitation suggested (but not absolutely required) by 
the optical results, CH in the diffuse ISM could in principle be
inverted (perhaps fairly strongly) or not, and we should be prepared 
to encounter situations where the the same N(CH) can produce different 
microwave CH profile integrals and \Nmu(CH).  The usual practice of 
deriving \Nmu(CH) in the limit of weak inversion is not necessarily
appropriate in all diffuse gas.

%
%

\begin{figure*}
\psfig{figure=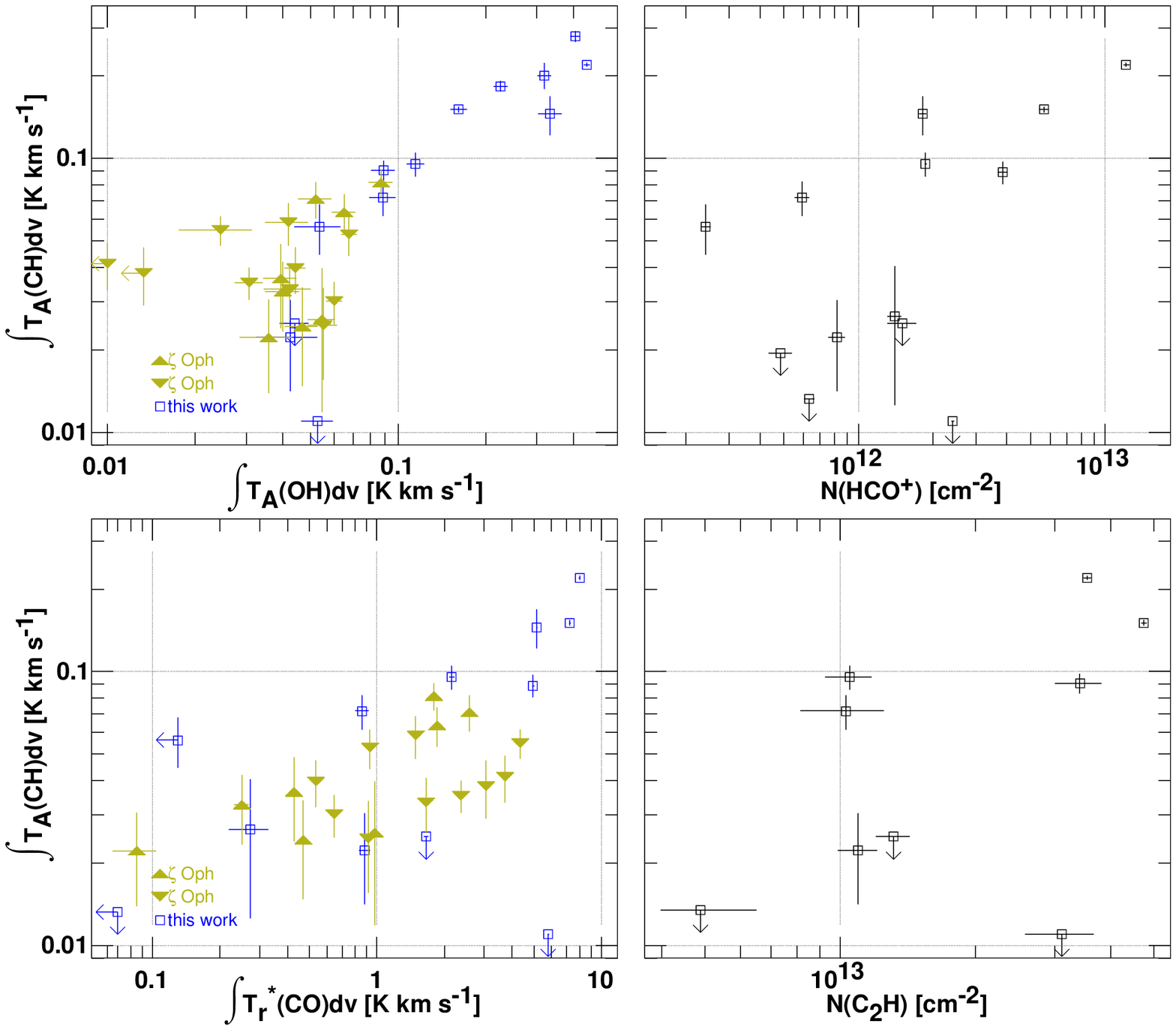,height=14.2cm}
\caption[]{Comparison of 3335 MHz F=1-1 CH profile integrals  with
other quantities. Left top: comparison with the OH profile integral 
toward compact continuum sources  \citep{LucLis96,LisLuc96,LisLuc00}
 and (heavily shaded) the data of  \cite{Lis97} around \zoph .  
Left bottom: the CO J=1-0 profile from our 12m telescope data 
\citep{LisLuc98} and (grayed) the CO data around
\zoph\ from \cite{Lis97} (see Sect. 2.3).  Right top: the \hcop\
column density derived from 89 GHz absorption \citep{LucLis96}.
Right bottom: the \cch\ column density from mm-wave absorption
\citep{LucLis00a}.}
\end{figure*}

\subsection{Abundances of hydrides in diffuse and dark gas}

Shown in Table 1 are the optically-determined column densities of the 
simple hydrides of C, N, and O in the only directions in which they may 
be compared.  Taking weighted averages, we have for the means and their 
formal variance -- taken over a very limited range of N(\HH) --
N(CH)/N(\HH), $4.4\pm1.0\times10^{-8}$;
N(OH)/N(\HH), $1.0\pm0.2\times10^{-7}$;
N(NH)/N(\HH), $1.9\pm0.1\times10^{-9}$; 
N(OH)/N(CH), $3.0\pm0.9$;
N(OH)/N(NH), $50.9\pm5.1$. 

Also shown are the microwave-determined column densities of OH and CH
toward TMC-1 and L134N, taken from \cite{OhiIrv+92}.  \XCH\ declines
markedly (a factor of 2-4) when carbon is in the form of CO rather 
than C$^+$, while X$_{\rm OH}$ remains constant or increases slightly.

\section{Microwave CH lines in diffuse gas}

\begin{table}
\caption[]{Background sources observed}
{
\begin{tabular}{lccc}
\hline
Source&l& b & \Tsub C $^1$ \\ 
\hline
B0212+735 & 128.93\degr & 11.96\degr & 0.79 K \\
B0224+671 & 132.12\degr & 6.23 \degr & 0.56 K \\
B0355+508 & 150.38\degr  & $-$1.60\degr  & 0.90 K \\
B0415+379 & 161.68\degr & $-$8.82\degr  & 2.79 K \\
B0433+295 & 170.58\degr & $-$11.66\degr  & 7.87 K \\
B0459+252 & 177.73\degr & $-$9.91\degr  & 1.02 K \\
B0528+134 &  191.37\degr &$-$11.01\degr  & 0.75 K \\
B0727-115 & 227.77\degr& 3.14\degr  & 0.91 K \\
B0736+017 & 216.99\degr & 11.38\degr  & 0.68 K \\
B0954+658 & 145.76\degr & 43.13\degr  & \\
B1954+513 & 85.20\degr & 11.90\degr  & 0.44 K  \\
B2013+370 & 74.77\degr & 1.36\degr  & 1.95 K \\
B2023+336 & 73.03\degr & $-$2.23\degr  & 0.85 K \\
B2200+420 &  92.13\degr &$-$10.40\degr  & 1.69 K \\
\hline
\end{tabular}}
\\
$^1$ gain was approximately 2.5 Jy K$^{-1}$ \\
\end{table}

\subsection{Data and new 3335 MHz observations}

The new 3335 MHz F=1-1 CH observations discussed here were taken with the 
now-defunct NRAO 43m antenna in 1995 September.  The data taken toward \zoph\ 
were reported previously in  \cite{Lis97}.  The continuum background sources 
observed are given in Table 2;  with the exceptions of 3C123 (B0433+295) and 
3C133 (B0459+252) they are all strong mm-wave sources.  The observing strategy 
was like that followed for OH  \citep{LisLuc96} and \hhco\  \citep{LisLuc95a} 
whereby we observed toward the source and 1.2 HPBW (here, 
$1.2 \times $9\arcmin) displaced in the four cardinal directions.  The 
beam efficiency of the telescope $\eta_B$ is usually quoted as 0.7-0.8.  We
take  $\eta_B = 0.7 $ because on this scale our profile toward L134N
reproduces the canonical value for \Nmu(CH) cited in Table 1.
We used a channel  separation of 2.44 kHz or 0.22 \kms.  The off-source 
system temperature was typically 38-42K.

The relationship between brightness of the microwave lines and CH 
column is nearly always taken in the Rayleigh-Jeans limit for small 
optical depth because 3335 MHz corresponds to 0.160 K and the excitation
temperature is usually taken as either -15 K, following the original 
observation of Perseus and Orion Arm features toward Cas A 
\citep{RydKol+76,HjaSum+77} or -60 K  \citep{GenDow+79}.  These values 
imply that the optical depth is very low, because typical brightnesses are 
0.05 K.  From atomic physics
and thermodynamic equilbrium we have for the 
$\nu = 3335~{\rm MHz}~F^\prime=1 \rightarrow F^{\prime\prime}=1$ 
transition the relationship

$$\int\tau_{11}dv = 
 {{0.75 {\rm N}_l \lambda^3 A_{11} (1-\exp{(-h\nu/k\Texc)})}\over{8\pi}} 
 \eqno(1)$$

where $\lambda = c/\nu, A_{11} = 2.04\times10^{-10}~\ps$ and
N$_l = {\rm N(CH)}/(1+\exp{(-h\nu/k\Texc)})$ is the population in the
lower half of the ground-state $\Lambda$-doublet, of which a fraction
3/4 is in the F$^{\prime\prime} = 1$ level.

This may be used in concert with the relationship between the observed
microwave antenna temperature \TA , the excitation temperature \Texc\ and 
the background continuum temperature against which the line is observed
\Tbg\ 

$$\TA = \eta_B \omet [J(\Texc)-J(\Tbg)] \eqno(2) $$

where $J(x) \equiv (h\nu/k)/(\exp{(h\nu/kx)}-1)$.  In the Rayleigh-Jeans
limit of small optical depth, with N(CH)/$(1+\exp{(-h\nu/k\Texc)}) \approx$ 
N(CH)/2 one arrives at the relationship

$$\Nmu(CH) = 
 {{2.82\times10^{14}\pcc \Texc \int{\TA dv}}\over{\eta_B(\Texc-\Tbg)}}
  \eqno(3) $$
In the literature, the mantissa of the constant on the right hand side 
of this expression is sometimes quoted as 2.9, rather than 2.82. 

Mean off-source spectra for many lines of sight are shown in Fig. 4 along 
with \hcop\ absorption data \citep{LucLis96} where possible.  Except in
two cases (see Sect. 3.4) the on and off-source CH spectra do not
differ significantly and the latter have lower noise levels since
they are four-point averages.  For \bll\ (B2200+420) we detected CH 
only toward the source as was also the case 
in our search for \hcop\ {\it emission}. Also shown are a comparison 
spectrum toward L134N as well as the mean of the nine spectra observed 
around \zoph\ and shown individually in  \cite{Lis97}.  Line profile
integrals and fitted gaussian components are given in the Appendix.
CH spectra toward two sources are compared with OH, \hcop\ and CO 
emission and absorption in Fig. 5.  

Fig. 4 shows that most but not
all \hcop\ absorption components are seen in CH emission: the
\hcop\ absorption feature toward B0212+735 which is missing in
CH emission is also absent in CO emission, but seen in CO
absorption \citep{LisLuc98}.  The absence of CH absorption hinders our 
ability to find \Nmu(CH) for components of low column density.  Given the vast
difference in resolution, the differences in CH and \hcop\ profile shapes
 however interesting, are remarkably slight.

Fig. 5 toward B0355+508 shows that CH, like OH, has a diffuse 
distributed component and does not clearly distinguish individual 
features in all cases.  As discussed by  \cite{LisLuc00}, \hcop\
absorption extends out to -36 \kms, corresponding to the main
body of H I absorption in this direction.  

Toward 3C111 (B0415+379)
the CH is strongly inverted only in the component at higher velocity, 
which is more chemically complex and likely more nearly fully molecular 
(see the discussion in  \cite{LucLis98}).  Gas in the lower-velocity 
component is heavily fractionated in carbon, indicating that only a 
small fraction of the carbon is in CO.  Note that the CO absorption 
profiles at bottom are for the CO J=2-1 line while the CO emission 
data shown at top are for J=1-0.  For a discussion of CO along this 
line of sight, see \cite{LucLis98} and \cite{LisLuc98}.

\subsection{CH, OH, \hcop and \cch}

Figure 6 compares the integrated (off-source) CH line strength with several 
other measures; the quantities used in the Figure are given in Table A.1 
in the Appendix. The \hcop\ and \cch\ absorption \citep{LucLis96,LucLis00a} 
and CO emission \citep{LisLuc98}, were taken {\it on-source} while the OH 
emission for continuum sources \citep{LisLuc96,LisLuc00} is also off-source 
data, but displaced further owing to the lower resolution.  In this Figure we 
have also included (heavily greyed) our CH, OH and CO data for the 
diffuse gas around \zoph\  \citep{Lis97}.  Shown as shaded upward 
(downward) triangles are the values observed at or to the North (South) 
of the star; the profile integrals for positive and negative velocity 
gas are shown separately because there are actually two components at 
each position.   Around \zoph, the observations of CH, 
OH and CO were concentric.  For purposes of nominal comparison 
\Nmu(CH) $\approx 4\times10^{14}\pcc \int{\TA dv}$ corresponding to
$\Texc << 0, \eta_B = 0.7$.

The OH and CH emission profile integrals are tightly correlated for the 
stronger components.   For these features the mean signal-noise weighted 
ratio of the CH and OH line profile integrals (in the sense CH:OH) is 
$0.67\pm0.19$.  Deriving \Nmu(CH) as given just above and  
taking the usual formulae for \Nmu(OH) ($e.g.$ Eqn. 3 of \cite{LisLuc96})
with \Texc-\Tcmb = 1 or 2 K for OH ({\it ibid}),  we have that 
\Nmu(OH)/\Nmu(CH) = 4.3 or 2.8, in agreement with the result derived 
optically and quoted in Sect. 2.4,
N(OH)/N(CH) = $3\pm0.9$. This serves as a helpful confirmation of the
unexpectedly rather small OH main-line excitation temperatures
seen in diffuse gas in the radio and optical \citep{Rou96} regimes.
Indirectly, it also suggests that the CH is only weakly inverted
($|\Texc|$ is large).

There appear to be two branches of the CH/OH ratio for weak features,
one in which the CH brightness increases fairly abruptly at 
$\int{\TA({\rm OH})dv} = 0.04-0.05$ K \kms and another (seen South of \zoph) 
where the CH is relatively strong at small values of the OH integral.  The 
weak CH features showing complicated behaviour in OH and CH are double-valued 
but well-ordered with respect to a comparison of CH and CO as discussed in 
Sect. 3.3.


The comparison with both \hcop\ and \cch\ absorption seems to show 
some sort of bimodality or, at least, a very large range of CH profile 
integrals at a given N(\hcop) or N(\cch).   If the fractional abundances 
of CH and \hcop\ are both about constant (which is otherwise believed to
be the case for diffuse gas) the 3335 MHz profile integral (\Nmu(CH)) 
does not have a single proportionality to N(CH) for N(\hcop) 
$< 3\times10^{12}\pcc$. Taking the data at face value, for N(\hcop) 
$ = 10^{12}\pcc$ we have that \Nmu(CH) $ \approx 0.8-3\times10^{13}\pcc$ 
or 
N(\HH) $= 2 - 8 \times 10^{20}\pcc$, leading to 
N(\hcop)/N(\HH) $ = 1.25 - 5 \times 10^{-9}$.
The same comparison at  
N(\hcop) $ = 3\times 10^{12}\pcc$ and
\Nmu(CH) $ \approx 5\times10^{13}\pcc$
yields N(\hcop)/N(\HH) $ = 2.5 \times 10^{-9}$.
Earlier, we inferred 
N(\hcop)/N(\HH) $ = 2-3 \times 10^{-9}$ from the
observed tight relationship between \hcop\ and OH
\citep{LucLis96,LisLuc96} and we showed that such
an unexpectedly high \hcop\ abundances suffices to
explain the observed quantities of CO for all
N(CO) $\la 3\times 10^{16}\pcc$ \citep{LisLuc00}.
A cross-comparison of the N(CN)/N(\HH) ratios determined 
optically with the column densities of CN and \hcop\
seen at mm-wavelengths by \cite{LisLuc01} also confirmed
this abundance of \hcop.

\subsection{CH and CO}

For the comparison with CO emission at lower left in Fig. 6, the new data 
toward continuum sources seemed also to indicate bimodality for the weaker 
emission.  The combined dataset including the points taken around 
\zoph\ clearly confirms this behaviour for intermediate values of the 
CO emission profile integral 0.3 $\le$ W(CO) $\le$ 5 K \kms.  The CO 
emission profile integral is definitely 
not bimodal with respect to either N(\hcop) or N(CO) 
over this range \citep{LisLuc98}.  A direct comparison between 
the CO column density measured in absorption and the mm-wave 
emission brightness  shows that 
N(CO) $ \approx 1.25 \times 10^{15}\pcc \int{\Tstar r dv}$ for
$ \int{\Tstar r dv} = 0.3 - 4 $ K \kms, see Fig. 12 of
\cite{LisLuc98}.


When the CH brightness is assumed to be proportional to the
CH column density, the correlation between CH and \HH\ then
provides for a CO brightness - \HH\ column density conversion
as well.  \cite{MagOne95} recently compared brightnesses of CH and CO, 
deriving N(\HH) from \Nmu(CH) by use of the CH-\HH\ conversion of \cite{Mat86}
as shown in Fig. 1 and discussed in Sect. 2.1 here.  \cite{MagOne95} noted a 
lot of scatter in the CO brightness - \HH\ column density conversion factors 
so derived, but did not plot the underlying data, which appear in
graphical form here in Fig. 7 (nor did they note the bimodal behaviour
which is clearly seen there).  In passing we note that much of the data for 
so-called translucent gas was considered to sample diffuse material in the 
original reference \citep{FedWil82}.

The chained lines (connected solid symbols) in Fig. 7 show the CH and CO data 
tabulated by \cite{MagOne95}, manipulated only to the extent that one source 
considered translucent (3C353) has been reclassified as dark since an examination 
of the original reference \citep{FedEva+87} showed that a visual extinction of 
6 magnitudes was quoted for the position whose \Nmu(CH) was used.
The data from our Fig. 6 at lower left have been transferred to Fig. 7, and 
it is apparent that they overlap the ``translucent'' data of \cite{MagOne95} 
except at the very highest values of the CO profile integral.  Also shown in 
Fig. 7 is the more recent mapping data of \cite{MagOne+98} for two high-latitude 
clouds described as translucent.  The observations for one of them, MBM40, all 
fall into the ``translucent'' regime; data for  MBM 16 lie chiefly on the 
``dark'' gas locus, except at the lowest values of W(CO), and could be 
interpreted as showing an abrupt transition from diffuse to dark conditions at 
W(CO) = 1 K \kms.

We can distinguish three modes of behaviour for the data in this Figure.  
First, there is a slow increase of \Nmu(CH) with W(CO) for both the dark
and diffuse/translucent gases: possible appearances aside, the best-fit 
power-law slope for the ``dark'' data of \cite{MagOne95} has 
\Nmu(CH) $\propto$ W(CO)$^{0.34\pm0.05}$ while 
\Nmu(CH) $\propto$ W(CO)$^{0.26\pm0.12}$ for the data along the line
marked as ``translucent.''  
Second there is a nearly fixed, factor of 3 
offset in \Nmu(CH) between the dark cloud and diffuse/translucent material.  The 
CH brightness appears bimodal at a given value of the CO profile integral 
in a variety of separate datasets and the offset in \Nmu(CH) persists up to 
CO profile integrals at least as large as 20 K \kms, which represents quite a 
strong CO line.  For very small CO profile integrals W(CO) $<$ 1 K \kms, the 
dark and diffuse regimes may merge, which is intuitively understandable.
The third discernible mode of behaviour is scatter in \Nmu(CH) at a fixed
W(CO) which we ascribed above to variations in CH excitation and other 
local conditions, for the diffuse/translucent gas.

The strength of CO emission is expected to change rapidly in diffuse 
gas: N(\HH) $\propto \EBV^2$ from Fig. 1, and 
N(CO) $\propto$ N(CH)$^2 \propto$ N(\HH)$^2$ 
\citep{FedLam88,FedStr+94,LisLuc00}.  So, it follows that the CH profile
integral should vary much less rapidly than the CO brightness if
\Nmu(CH) $\propto$ N(CH). The change 
in N(\HH) across Fig. 7 should be smaller than that in W(CO), for the 
diffuse gas at least, which would lead to a rather shallow slope.
But it is far from obvious that the diffuse/translucent and dark gas components 
should have so nearly the same slope in their
\Nmu(CH)-W(CO) relationship.  Perhaps the shallowness of the slope in the
dark gas reflects the decline of \XCH\ in denser gas, at higher W(CO).

A literal interpretation 
of Fig. 7 implies that dark gas requires about 3 times as much N(\HH) to produce 
a given W(CO) for W(CO) = 1-20 K \kms\ (this difference might be larger if
\Nmu(CH) is diminished in dark gas because \XCH\ is smaller).  If such were
the case, a substantial CO emission component from diffuse gas mixed into
an ensemble of dark clouds could cause misestimation of N(H). However, there is 
no other evidence for a diminished
N(\HH)/W(CO) ratio in diffuse gas, and actual measurements, in
the few possible cases, yield a typical value.  For instance, from
a comparison of the CO and \hcop\ data in Figs. 12 and 13 of \cite{LisLuc98}, it 
follows that N($\HH$)/W(CO) $\approx 2.5\times10^{20}~\HH~\pcc$/(K \kms)
around W(CO) = 1-2 K \kms.  Toward \zoph, where all the relevant
properties are directly measureable, we have
N($\HH$)/W(CO) $\approx 4\times10^{20}~\HH~\pcc$/(K \kms) \citep{Lis82,Lis97}.

\begin{figure}
\psfig{figure=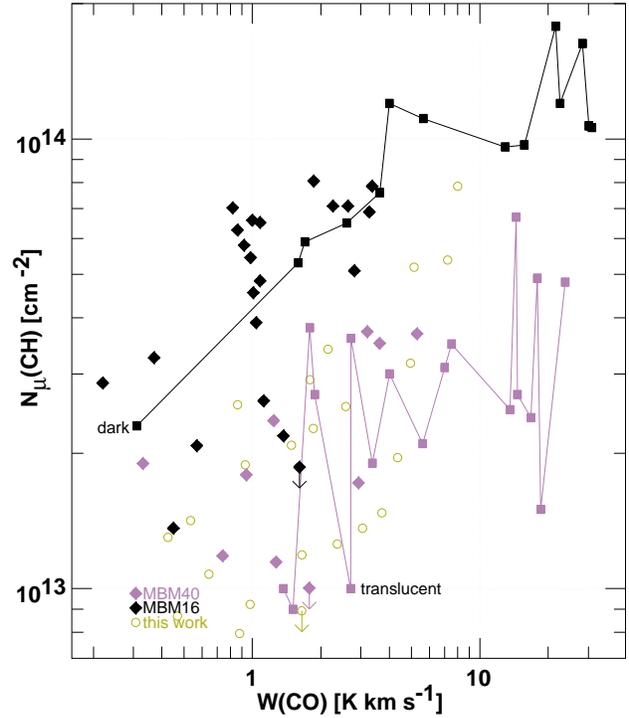,height=9.5cm}
\caption[]{Microwave-derived CH column densities $\Nmu$(CH) and CO 
profile integrals W(CO).   The solid, connected symbols represent data 
tabulated by \cite{MagOne95} for lines of sight considered 
\lq translucent \rq (shaded) and \lq dark \rq (shown dark).   Mapping data for 
two high-latitude  molecular clouds from \cite{MagOne+98} are shown 
as dark and shaded solid diamonds.  Our data from Fig. 6 at 
bottom left have been overlaid as the open circles.}
\end{figure}

\begin{figure}
\psfig{figure=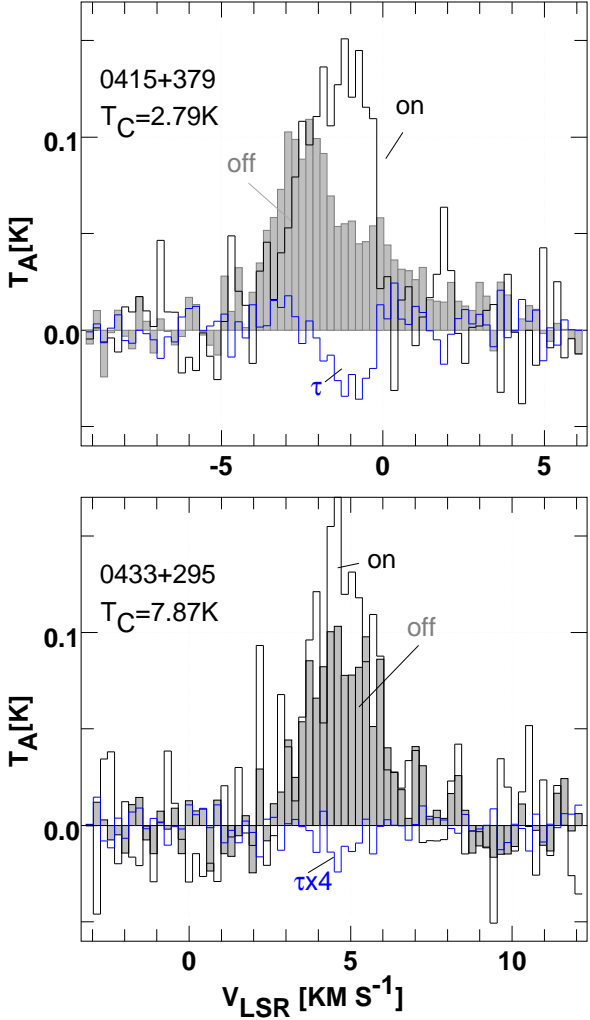,height=13.5cm}
\caption[]{On and off source spectra toward two strong
continuum sources.  The off-source average is shown shaded.
An optical depth spectrum derived from Eqn. 2 is shown in
each case; for B0433+295 (3C123) $\Texc = -10.7\pm3.2$ for the 
five contiguous central channels showing $\tau < 0$.}
\end{figure}



\subsection{Excitation}

Notably, the on-source CH profile is not weaker than the nearby 
off-source average in a single direction out of all those observed;
the CH $\Lambda$-doublet is generally inverted in diffuse
gas.  Fig. 8 shows on and off-source spectra taken toward 3C111 and
3C123, the strongest continuum sources observed.  Also shown
are optical depth spectra made by applying Eqn. 2 to the
data taken on ($\Tbg = \Tcmb+\Tsub C $) and off-source ($\Tbg = \Tcmb$).
In only one case (3C123) can we derive a statistically significant 
inversion ($ \Texc < 0 $).  Even for 3C111, the excitation temperature 
derived from a further application of Eqn. 2 (using the derived 
optical depth profile) is positive even where $\tau < 0$ in Fig. 8; 
spatial structure in the CH emission renders the solutions 
unreliable.

For 3C123, however, $<$\Texc$>$  = -10.7$\pm$3.2 K averaged over the 
five contiguous central channels where $\tau < 0$  in Fig. 8 and 
$-16 < \Texc < -6$ K over this same interval.  This is essentially
the same result as that obtained by \cite{HjaSum+77}, 
$\Texc = -9\pm5$ K. \cite{GenDow+79} quote a value 
$\Texc = -60\pm30$ K in this direction but the meaning of the 
(rather large) error or range  is not clear from their presentation.  
\cite{GenDow+79} thus claim that the 3335 MHz transition seen with the 
100m telescope is substantially less inverted than was found earlier 
using the Onsala 25m dish.
In the optically thin, Rayleigh-Jeans limit of the column density 
determination  \Nmu(CH) $\propto \Texc/(\Texc-\Tcmb)$.  The
\Texc-dependent term in this expression is 0.96 for $\Texc = -60$ K,
0.85 for $\Texc = -15$ K, and  0.79 for $\Texc = -10$ K.

\section{Summary}

Although fourth in the current series, this paper is actually also the third 
of three papers discussing singledish cm-wave spectra of the molecules in 
diffuse gas, taken toward and around a sample of compact, extragalactic 
mm-wave continuum sources.  In H I, such on-off comparison experiments have 
come to be known as emission-absorption experiments \citep{DicTer+78} but as 
it turns out, we performed one emission-absorption experiment (in OH; 
\cite{LisLuc96}), one absorption-absorption experiment (in \HH CO, which 
appeared in absorption both on and off-source; \cite{LisLuc95a}) and,
here, one emission-emission experiment (since no true absorption was
detectable).

We began by displaying the richly structured behaviour of CH with \EBV;
N(CH) is multi-valued with respect to \EBV, depending on the degree of
conversion to molecular gas along the line of sight, and a simple, linear  
CH-\EBV\ relationship can be expected only when the extinction is dominated 
by molecular gas, as toward a single dark cloud.  Otherwise, the range of
measured N(CH) at a given \EBV\ in the diffuse gas as a whole is typically
more than one order of magnitude.  Much of this behaviour can be explained
on the basis of an easily-demonstrated and long-known, nearly constant
relative abundance $<$\XCH$>$ = $4.3\pm1.9\times10^{-8}$, and 
N(CH) $\propto$ N(\HH)$^{1.00\pm0.06}$ for N(\HH) $\la 10^{21}~\pcc$: as well,
we have that N(\HH) $\propto$ \EBV$^{1.8}$ for 0.2 $<$ \EBV\ $<$0.7. 

If CH is 
a good predictor of \HH, the 140 lines of sight gathered to study the 
CH-\EBV\ relationship allow derivation of the molecular fraction in the
diffuse/translucent ISM over a much wider range of sample mean densities 
$<$\EBV$>$/$<$R$>$ than is directly accessible in measurements of the
lines of hydrogen.  The molecular fraction found in this way is in
good agreement with direct measurements at 
low (Copernicus) and high (FUSE) sample mean density,
and is 0.4-0.45 for $<$\EBV$>$/$<$R$> =  0.61$ mag kpc$^{-1}$,
which is the accepted mean in the gas within 500-1000 pc.

We pointed out that sensitive optical measurements of the population 
ratio in the upper and lower halves of the ground-state CH 
$\lambda$-doublet toward two stars predict that the brightness of 
the microwave CH lines should be double-valued at a given CH column
density in diffuse gas depending on whether the excitation is inverted 
(the brighter branch); this is consistent with models of CH excitation 
which predict a transition from normal excitation to inversion at 
hydrogen densites in the range 10 - 1000 $\pcc$, but the effect is not 
present in the microwave lines in these directions.  This could be due 
to the disparity in beam-sizes or to relatively small errors in the
optical data.

We presented 3335 MHz CH observations toward some of the compact 
extragalactic mm-wave continuum sources studied in this series of papers, 
toward two strong cm-wave sources, and around \zoph, and compared the 
properties of CH with those of OH and CO in emission and \hcop\ and \cch\ 
seen in absorption. In stronger-lined gas, the CH/OH comparison confirms the 
very small OH excitation temperatures which have been found in
diffuse gas.  Comparisons of CH with \hcop\ and \cch\ show that
there is either a very large scatter in the CH brightness or 
microwave-derived CH column density at a given N(\hcop) or N(\cch) 
or perhaps a bimodality.  The CH/\hcop\ comparison readily (but
only roughly) confirms our previously-derived ratio 
N(\hcop)/N(\HH) $= 2 \times 10^{-9}$.

The 3335 MHz line brightness in diffuse gas is very definitely bimodal
with regard to CO emission in both the new data presented here and
our previously-published data around \zoph.  To explore this further,
we compared our data with other published studies which used CH to derive 
the CO-\HH\ conversion factor in diffuse/translucent gas and found that they 
are entirely consistent with our data.  This may be a manifestation of
a disparity between inverted and non-inverted CH expected in diffuse gas.

We found some hitherto-unnoticed systematic
behaviour in the CH-CO comparison in diffuse and dark gas, in particular a 
steady, factor of $\approx 3$ offset in the ratio of CH and CO profile 
integrals for W(CO) = 1 - 30 K \kms:  W(CH)/W(CO) is consistently larger 
by $\approx 3$ in dark gas. A shallow slope in the W(CH)-W(CO) relationship 
in diffuse gas is 
undertandable because the CO abundance varies rapidly with N(\HH), 
N(CO) $\propto $ N(\HH)$^2$ and the CO brightness will increase even
faster than N(CO), but the presence of nearly the same shallow slope 
W(CH) $\propto W({\rm CO})^{0.3}$ in dark and diffuse gas is puzzling.  
It may reflect the decline of \XCH\ which is known to occur in very dark
gas.

The next paper in this series will discuss several species whose 
abundances are best determined at cm-wave frequencies, such as \ammon, 
\hhco\ and C$_4$H.

\begin{acknowledgements}

The National Radio Astronomy Observatory is operated by AUI, Inc. under a
cooperative agreement with the US National Science Foundation.  IRAM is
operated by CNRS (France), the MPG (Germany) and the IGN (Spain). 
The comments of the referee, John Black, were very helpful.

\end{acknowledgements}

\bibliographystyle{apj}
\bibliography{mnemonic,absorption}

\begin{appendix}

\section{Line profile integrals shown in Fig. 6}

\begin{table*}
\caption[]{Line profile integrals}
{
\begin{tabular}{lcccccc}
\hline
Source & V & W(OH) & $\int{\tau(\hcop)dv}^1$ &  
 $\int{\tau(\cch)dv}^2$ & W(CO) & W(CH) \\
 & \kms & K \kms & \kms & \kms & K \kms & K \kms \\
\hline
B0212+735 &-10 & 0.0886(0.009)& 0.58(0.04) & 0.38(0.18) &0.86(0.06)& 0.0719(0.01033) \\
B0212+735 & 0 & $<$0.0134 & 0.62(0.03) & 0.18(0.06)& $<$0.07 & $<$0.01342 \\ 
B0212+735 & 3 & 0.0889(0.0084) &3.77(0.17) & 1.26(0.15)& 4.96(0.06) & 0.0906(0.00748) \\
B0415+379$^3$ & -2.2 &0.4443(0.0106) &11.91(0.334)& 1.31(0.03)& 8.00(0.07)& 0.2200(0.00389)  \\
B0415+379$^4$ &  -0.8 & 0.1614(0.0105) &5.526(0.244)& 1.75(0.02) &7.22(0.06)& 0.1506(0.0037) \\
B0433+295 & all & 0.4073(0.0157) &&&& 0.2792(0.01294) \\
B0459+252 & all & 0.2251(0.0126) &&&& 0.1835(0.00838) \\
B0528+134 & 2   & 0.0534(0.0098)&0.2362(0.0103)&&$<$0.13& 0.0563(0.01171)  \\
B0528+134 & 10  & 0.1146(0.0097)&1.836(0.019)&0.39(0.05) & 2.15(0.06)&   0.0955(0.00961) \\
B0727-115 & all & &0.476(0.052) &&& $<$0.0191 \\
B0736+017 & all& 0.0426(0.0101) &0.802(0.063)& 0.40(0.04)&0.88(0.05)&  0.0223(0.00813) \\
B0954+658 & all & 0.0442(0.0050) &1.48(0.20) &0.48(0.04)& 1.65(0.04)& $<$0.025 \\
B2013+370 & all & 0.3332(0.0310) &1.785(0.0236) &&5.15(0.23)& 0.1452(0.02359) \\
B2023+336 & all & 0.3185(0.0173) &&&&  0.2009(0.0219) \\
B2200+420 & all & 0.0529(0.0066) &2.36(0.03) &1.15(0.20) &5.78(0.05) & $<$0.011 \\
\hline
\end{tabular}}
\\
$^1$ N(\hcop) $= 1.02\times10^{12}\pcc \int{\tau(\hcop)dv}$ \\
$^2$ N(\cch) $= 2.711\times10^{13}\pcc \int{\tau(\cch)dv}$ \\
$^3$ \hcop\ data toward B0415+379 are 60$\times$ N(H$^{13}$CO$^+$) \\ 
$^4$ Toward 3C111 components were separated by Gaussian fitting \\
\end{table*}

\end{appendix}

\end{document}